\documentclass[preprint, 3p, twocolumn]{elsarticle}
\usepackage{graphicx, bm, epsfig, hyperref}
\biboptions{super, sort&compress, round}
\begin{document}
\title{Discrete Transformations in the Thomson Problem}
%\email[]{tim.lafavejr@utdallas.edu}
\author{Tim LaFave Jr.}%\corref{cor1}\fnref{fn1}
%\cortext[cor1]{tim.lafavejr@utdallas.edu}
%\fntext[fn1]{University of Texas at Dallas, Department of Electrical Engineering, Richardson, TX 75080}
%\affiliation{University of Texas at Dallas, Department of Electrical Engineering, Richardson, TX 75080}
\date{\today}
\begin{abstract}
A significantly lower upper limit to minimum energy solutions of the electrostatic Thomson Problem is reported. A point charge is introduced to the origin of each $N$-charge solution. This raises the total energy by $N$ as an upper limit to each $(N+1)$-charge solution. Minimization of energy to $U(N+1)$ is well fit with $-0.5518(3/2)\sqrt{N}+1/2$ for up to $N$=500. The energy distribution due to this displacement exhibits correspondences with shell-filling behavior in atomic systems. This work may aid development of more efficient and innovative numerical search algorithms to obtain $N$-charge configurations having global energy minima and yield new insights to atomic structure.
\end{abstract}
\begin{keyword}
``T. LaFave Jr. {\it J. Electrostatics} {\bf 72}(1) 39-43 (2014)''. DOI: 10.1016/j.elstat.2013.11.007
\end{keyword}

\maketitle
\section{Introduction}
The Thomson Problem has drawn considerable interest since the mid-1900s\cite{lafave-whyte1952} having found use in modeling fullerenes,\cite{lafave-garrido2000, lafave-kroto1985} drug encapsulants,\cite{lafave-espinoza1999} spherical viruses,\cite{lafave-caspar1962} and crystalline order on curved surfaces.\cite{lafave-bowick2002} In the past few decades several computational algorithms have yielded precise numerical solutions for many-$N$ electron systems.\cite{lafave-erber1991, lafave-glasser1992, lafave-edmundson1992, lafave-edmundson1993, lafave-altschuler1994, lafave-erber1995, lafave-PGarrido1996, lafave-erber1997, lafave-garrido1999, lafave-altschuler2005, lafave-altschuler2006, lafave-wales2006} The Thomson Problem has emerged as a benchmark for global optimization algorithms\cite{lafave-altschuler2005, lafave-altschuler2006} though its general solution remains unknown.\cite{lafave-smale1998} 

Numerical solutions of the Thomson Problem are those for which the total Coulomb repulsion energy,

\begin{eqnarray}\label{eq:lafave-thomson}
U(N) = \sum_{i<j}^N \frac{1}{\left| r_i - r_j \right| },
\end{eqnarray}

{\noindent}is a minimum for each $N$-charge system with $r_i$ and $r_j$ constrained to the surface of a unit sphere. 

The initial condition of some minimization algorithms is the random distribution of $N$ point charges on the unit sphere. This sets a relatively distant upper energy limit, $U_r(N) = N(N-1)/2$,\cite{lafave-glasser1992} from which numerous iterations progress toward a global minimum for each $N$-charge system. The distribution of numerical solutions has been fit with empirical functions including $U(N) = N^2/2 + aN^{3/2}$\cite{lafave-erber1991} and $U(N) = N^2/2 + aN^{3/2} + bN^{1/2}$.\cite{lafave-glasser1992} The quadratic term is ascribed to energy stored within a continuous charged shell of unit radius having total charge, $N$. With this interpretation, the half-integer terms may correspond to self-energies of $N$ uniformly charged disks that are removed to yield the final minimized energy of discrete charges. If the $N^2/2$ term is associated with the random distribution of point charges, the half-integer terms may be related to correlation energies of surface Coulomb equilibrium states.\cite{lafave-erber1991} These ascriptions of energy terms to physical entities have guided the development of fairly useful minimization algorithms.

Here, the discrete derivative of $N^2/2$ is shown to correspond to the introduction of a single point charge, $q_0$, at the origin of a given $N$-charge solution, and the discrete derivative of the remaining half-integer term(s) accounts for energy needed to displace $q_0$ to the unit sphere. This yields each subsequent $(N+1)$-charge solution of the Thomson Problem. In this manner, the upper limit of each minimized $(N+1)$ energy solution is given by $U(N)+N$. This represents a well-defined charge configuration with a significantly lower energy than random charge distributions and may be useful as initial conditions in relatively more efficient energy minimization algorithms.

Notably, the distribution of energy solutions of the Thomson problem is ``systematic''\cite{lafave-glasser1992} and not random. Here, having ignored the linear term in the discrete derivative, the remaining ``systematic'' distribution of energy associated with the displacement of $q_0$ to the surface of the unit sphere demonstrably exhibits features uniquely correspondent with numerous shell-filling features throughout the periodic table of natural atomic systems.\cite{lafave-lafave2006, lafave-lafave2008, lafave-lafave2013} Hence, introduction of $q_0$ to the Thomson Problem represents an intriguingly fresh perspective of this extensively useful mathematical model and the classical underpinnings of atomic structure.

Numerical solutions of the Thomson Problem for up to 500 charges used in this communication have been obtained from a continually updated database maintained by Syracuse University.\cite{lafave-thomsonApplet}

\section{Discrete Derivatives}
Two upper energy bounds of some interest are those associated with a continuous charge shell\cite{lafave-glasser1992} consisting of infinite-many $N$ charges such that

\begin{eqnarray}\label{eq:lafave-shell}
U_{\infty}(N) = \frac{N^2}{2}
\end{eqnarray}

{\noindent}and the random distribution of $N$ discrete point charges across the surface of the unit sphere

\begin{eqnarray}\label{eq:lafave-random}
U_r(N) = \frac{N}{2}(N-1) = \frac{N^2}{2}-\frac{N}{2}.
\end{eqnarray}

\begin{figure}%
\begin{center}
\includegraphics[width=7.5cm]{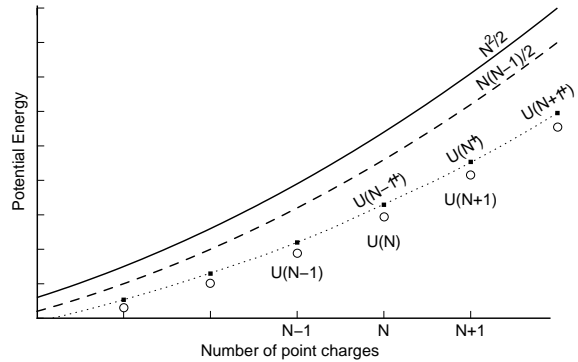}%
\end{center}
\caption{The extreme upper energy limit of the Thomson Problem is given by $N^2/2$ for a continuous charge shell followed by $N(N-1)/2$, the energy associated with a random distribution of $N$ point charges. Significantly lower, $U(N^+)$, the energy of a given $N$-charge solution of the Thomson Problem with one charge at its origin is readily obtained by $U(N)+N$, where $U(N)$ are solutions of the Thomson Problem.}%
\label{fig:lafave-Uplus}%
\end{figure}

{\noindent}Plots of Eqs.~\ref{eq:lafave-shell} and \ref{eq:lafave-random} are shown in Fig.~\ref{fig:lafave-Uplus} together with a few numerical solutions (open circles) of the Thomson Problem for illustration.

The minimized global potential energy solutions for up to $N$=65 point charges were previously fit using,\cite{lafave-erber1991}

\begin{eqnarray}\label{eq:lafave-erber}
f_1(N) = \frac{N^2}{2} + aN^{3/2}
\end{eqnarray}

{\noindent}in which $a=-0.5510$, and for up to $N\sim 100$ using\cite{lafave-glasser1992}

\begin{eqnarray}\label{eq:lafave-glasser}
f_2(N) = \frac{N^2}{2} + aN^{3/2} + bN^{1/2}.
\end{eqnarray}

These smooth, continuous fit functions are generally unrepresentative of the absolute minimum energy configurations of discrete charges due to a variety of issues. Among them, discrete charges cannot be infinitesimally subdivided so all points, $f(N)$, for non-integer $N$ have no physical significance without imposing additional arguments. If these intermediate values of $f(N)$ should have physical usefulness, for instance if an effective fractional charge on the unit sphere surface is admitted as a discrete charge approaches or leaves the surface, $f(N)$ should have local minima at integer values of $N$. Though these fit functions, Eqs.~\ref{eq:lafave-erber} and \ref{eq:lafave-glasser} have no local minima at integer values of $N$, they are coarsely instructive.

It is potentially more fruitful to design a fit function in accordance with the physical nature of electrostatic charge configurations. In particular, knowledge of $f(N)$ at integer values of $N$, given the discrete nature of point charges is paramount. Consider the discrete derivative of $f(N)$ at integer values of $N$, 

\begin{eqnarray}
\frac{\Delta f_i(N)}{\Delta N} = \frac{f(N+\Delta N) - f(N)}{\Delta N}
\end{eqnarray}

{\noindent}of Eq.~\ref{eq:lafave-erber}, which yields, after binomial expansion of the half-integer term,

%\begin{widetext}
\begin{eqnarray}\label{eq:lafave-erberprime}
\nonumber\!\!\!\!\!\!\!\!\!\!\Delta f_1(N) \!\!\!\!&=&\!\!\!\! \left(\!\!N + \frac{1}{2} \right)\\
&&\!\!\!\!+ \frac{a}{2}\!\!\left( \!\!3N^{\frac{1}{2}} + \frac{3}{4}N^{-\frac{1}{2}} - \frac{1}{8}N^{-\frac{3}{2}} + \cdots \right)
\end{eqnarray}
%\end{widetext}

{\noindent}for $\Delta N = 1$. For comparison, the derivative of Eq.~\ref{eq:lafave-glasser}, with binomial expansion of the half-integer terms,

%\begin{widetext}
\begin{eqnarray}\label{eq:lafave-glasserprime}
\nonumber\!\!\!\!\!\!\!\!\!\!\Delta f_2(N) \!\!\!\!&=&\!\!\!\!\left( \!\!N + \frac{1}{2} \right) + \frac{1}{2} \left[ \!3aN^{\frac{1}{2}} + \frac{7}{4} (a+b)N^{-\frac{1}{2}}\right.\\
&& \left.  + \frac{1}{8}(a-b)N^{-\frac{3}{2}}+ \cdots \right]
\end{eqnarray}
%\end{widetext}

{\noindent}for $\Delta N = 1$. The first set of parentheses in Eqs.~\ref{eq:lafave-erberprime} and \ref{eq:lafave-glasserprime} is the discrete derivative of $N^2/2$, the energy of a continuous charge shell, Eq.~\ref{eq:lafave-shell}.

\section{Discrete Transformation Energies}
\begin{figure}%
\begin{center}
\includegraphics[width=7.5cm]{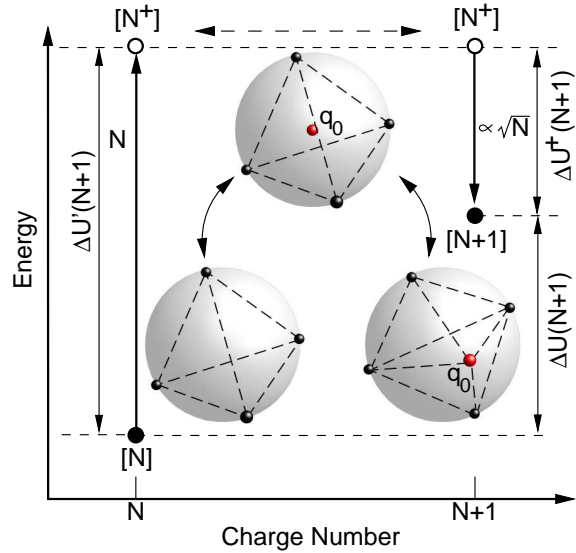}%
\caption{A well-defined intermediate charge configuration, $[N^+]$, with $q_0$ at the origin of $[N]$ linearly increases the energy by $N$. Consequently, $U(N^+)$ is readily accessible. Displacement of $q_0$ to the unit sphere surface and global minimization of energy yields the new $[N+1]$ solution with $\Delta U^+(N+1) \propto \sqrt{N}$.}%
\label{fig:lafave-symmetryDiagram}%
\end{center}
\end{figure}

To understand the physical charge distribution represented by the discrete energy differences, Eqs.~\ref{eq:lafave-erberprime} and \ref{eq:lafave-glasserprime}, consider an $N$-charge solution, such as that of $[N]$ shown in Fig.~\ref{fig:lafave-symmetryDiagram} for which the total minimized electrostatic energy is $U(N)$ as shown in Fig.~\ref{fig:lafave-Uplus}. The discrete energy changes needed to obtain the solution $U(N+1)$ in Fig.~\ref{fig:lafave-Uplus} may be obtained by introducing an $(N+1)$th point charge, $q_0$, to the origin denoted $[N^+]$ in Fig.~\ref{fig:lafave-symmetryDiagram}. Its contribution to the total energy, given equal interaction with all $N$ charges on the unit sphere, is $N$. The total energy, $U(N^+)$, (Fig.~\ref{fig:lafave-Uplus}) is $U(N)+N$ as illustrated in Fig.~\ref{fig:lafave-symmetryDiagram}. Thus, the first term of each discrete derivative, $\Delta f_i(N)$, is due to the appearance of $q_0$ at the origin. The remaining terms are due to the displacement of $q_0$ to the unit sphere surface and subsequent (or in situ) global minimization of energy, $U(N+1)$ in Fig.~\ref{fig:lafave-Uplus}, to yield the $(N+1)$ configuration (Fig.~\ref{fig:lafave-symmetryDiagram}).

In the strictest spirit of the Thomson Problem, introduction of $q_0$ to the {\it origin} of the unit sphere (Fig.~\ref{fig:lafave-symmetryDiagram}b) increases the total energy of charges on the unit sphere by $N/2$. However, since the goal is to incorporate $q_0$ in the final $(N+1)$-charge configuration on the unit sphere, inclusion of the energy associated with $q_0$, that is $N/2$, while at the origin is necessary. 

The example $[N^+]$ solution shown in Fig.~\ref{fig:lafave-symmetryDiagram}, consists of four charges at the vertices of a regular tetrahedron centered about $q_0$ at the origin. Symmetrically, the $[N^+]$ configuration is identical to the $[N]$ solution of the Thomson problem. In general, $U(N^+) > U(N+1)$. Therefore, the energy difference $\Delta U(N+1)$ associated with the addition of a charge in the Thomson problem may be expressed,

\begin{eqnarray}\label{eq:lafave-UN}
\Delta U(N+1) \!\!\!\!&=&\!\!\!\! \Delta U^{\prime}(N+1) - \Delta U^+(N+1)\nonumber\\
\!\!\!\!&=&\!\!\!\! N - \Delta U^+(N+1)
\end{eqnarray}

{\noindent}as shown in Fig.~\ref{fig:lafave-symmetryDiagram}. Consequently, $\Delta U^+(N+1)$ accounts for the half-integer terms in Eqs.~\ref{eq:lafave-erberprime} and \ref{eq:lafave-glasserprime} and the $+1/2$ term associated with the discrete derivative of $N^2/2$.

\begin{figure}%
\begin{center}
\includegraphics[width=7.5cm]{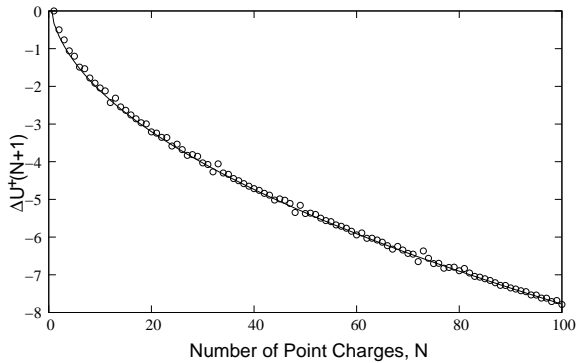}%
\caption{The function $\Delta f(N)=-0.5518(3/2)\sqrt{N}+1/2$, fit to $N$=500 numerical solutions of the Thomson Problem, is shown on the first $N$=100 numerical solutions of the energy difference $\Delta U^+(N+1)$.}%
\label{fig:lafave-fit}%
\end{center}
\end{figure}

To check the validity of the discrete transformations from each $N$-charge solution of the Thomson Problem to its adjacent $(N+1)$-charge solution, $\Delta U^+(N+1)$ in Eq.~\ref{eq:lafave-UN} is fit with the remaining terms of the discrete derivatives of Eqs.~\ref{eq:lafave-erberprime} and \ref{eq:lafave-glasserprime}. The distribution of energy differences for up to 500-charge numerical solutions of the Thomson Problem (Fig.~\ref{fig:lafave-fit}) is well-fit with the first three terms of Eq.~\ref{eq:lafave-erberprime} for which $a = 0.5512\pm 1\times 10^{-4}$, in very good agreement with previous work in which $a=0.5510$ for up to $N=65$.\cite{lafave-erber1991} Truncating Eq.~\ref{eq:lafave-erberprime} to only the first term, including the $+1/2$ term from the discrete derivative of the quadratic term in Eq.~\ref{eq:lafave-erber}, 

\begin{eqnarray}\label{eq:lafave-sqrt}
f^+(N+1) =\frac{3}{2}a \sqrt{N} + \frac{1}{2}
\end{eqnarray}

{\noindent}provides a reasonably good fit with $a=0.5518\pm 1\times 10^{-4}$. Fitting with half-integer terms in Eq.~\ref{eq:lafave-glasserprime} yields large errors associated with $b$ and larger values of $a = 0.5525$. These half-integer terms are, however, useful for better fits to smaller $N$-charge solutions.\cite{lafave-glasser1992} 

Since fitting smooth functions to a discrete, ``systematic'' data set is an insufficient means of characterizing solutions of the Thomson Problem, Eq.~\ref{eq:lafave-sqrt} is a convenient fit function for $\Delta U^+(N+1)$ for up to 500 charges as shown for up to 100 charges in Fig.~\ref{fig:lafave-fit}.

\section{The $q_0$ Transformation}

Applied to a continuous charge shell, introduction of $q_0$ (one of an infinite number of equal charge elements) to the origin raises the energy by $N$,

\begin{eqnarray}
U_{\infty}(N)\to U_{\infty}(N)+N = \frac{N^2+2N}{2}.
\end{eqnarray}

{\noindent}Here, the transformation yields a lower energy than $U_{\infty}(N+1)$,

\begin{eqnarray}
U_{\infty}(N) + N - U_{\infty}(N+1) = -\frac{1}{2}.
\end{eqnarray}

{\noindent}However, the energy difference of $-1/2$ accounts for the $+1/2$ term in the finite difference of $N^2/2$.

Applied to purely random charge distributions of energy given by Eq.~\ref{eq:lafave-random}, the $q_0$ transformation raises the energy of each $N$-charge configuration by $N$,

\begin{eqnarray}
U_r(N) \to U_{r}(N)+N = \frac{N^2+N}{2}.
\end{eqnarray}

{\noindent}Its difference from the energy of a random $(N+1)$-charge distribution is zero since 

\begin{eqnarray}
U_{r}(N+1) = \frac{(N+1)N}{2} = \frac{N^2+N}{2}.
\end{eqnarray}

{\noindent}Consequently, the $q_0$ transformation {\it generates} energies of purely random charge distributions. However, as there is no energy difference between $N+1$ randomly distributed charges on a unit sphere and $N$ charges on a unit sphere with $q_0$ at the origin, these two configurations are energetically indistinguishable. 

The $q_0$ transformation uniquely distinguishes between continuous and discrete charge systems. 
 
Applied to the Thomson Problem, the $q_0$ transformation raises the energy of each $N$-charge solution by $N$ and represents an energetically unfavorable charge configuration to the $N+1$ solution with all charges on the unit sphere as made evident by Fig.~\ref{fig:lafave-fit}. In more detail, using Eq.~\ref{eq:lafave-erber}, inclusion of $q_0$ to each $N$-charge solution imposes the following inequality on the coarse fit parameter, $a$,

\begin{eqnarray}
\!\!\!\!\!\!\!\!U(N)+N \!\!&>&\!\! U(N+1)\nonumber\\
aN^{3/2} \!\!&>&\!\! \frac{1}{2} + a(N+1)^{3/2}\nonumber\\
a \!\!&>&\!\! -\frac{1}{2} \left[ (N+1)^{3/2}-N^{3/2} \right]^{-1}.
\end{eqnarray}

{\noindent}For nonzero positive integer values of $N$, $a$ must be greater than $-0.914$ ($N=1$). Since the coarse fit parameter $a=-0.5518$ is greater than this threshold, the $q_0$ transformation yields an upper limit to all energy solutions of the Thomson Problem.
 
\section{Fit Functions}
Coarsely-defined fit functions $f_i(N)$ presume the possibility of fractional charges between integer values of $N$. Indeed, each $N$-charge solution represents a global electrostatic potential energy {\it minimum} that is not represented by these fit functions. Consequently, the underlying physical nature of discrete charge systems is in conflict with the mathematics used to study them. 

\begin{figure}%
\begin{center}
\includegraphics[width=7.5cm]{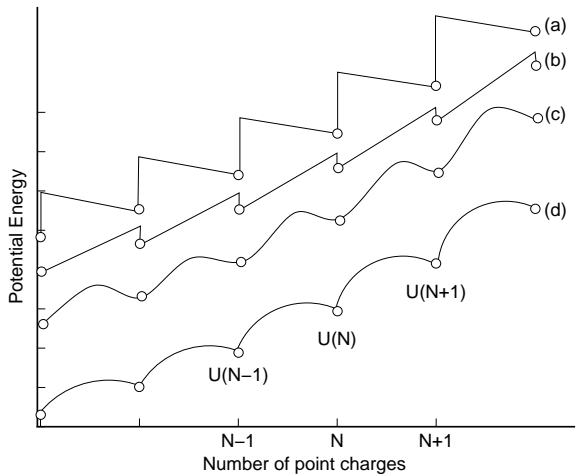}%
\caption{A few possible continuous fit functions to minimized energy solutions of the Thomson Problem. a) $q_0$ is introduced abruptly to each $N$-charge solution, and the energy is gradually minimized to $U(N+1)$. b) $q_0$ forms gradually at the origin, and the total energy is minimized to $U(N+1)$. c) Energy minima are obtained with a smooth, periodic function having minima at integer values of $N$. d) Energy solutions are obtained with a piecewise-continuous function with minima at integer $N$ values.}%
\label{fig:lafave-functions}%
\end{center}
\end{figure}

As a new point charge is added to a given $N$-charge configuration the $N$ charges respond to its arrival until an $[N\!+\!1]$ configuration is obtained. A fit function may be designed to reflect local minima at integer values of $N$. Consequently, the fit function may ``hop'' from one energy minimum to the next through discrete energy differences to satisfy the condition of local minima at integer values of $N$. A few possible fit functions are shown schematically in Fig.~\ref{fig:lafave-functions}. The saw-tooth fit function sketched in Fig.~\ref{fig:lafave-functions}a reflects abrupt introduction of $q_0$ at the origin of each $N$-charge configuration followed by a gradual relaxation of the function to the neighboring $[N\!+\!1]$ configuration. This function may be useful if the strict definition of charges on the {\it surface} of the unit sphere is to be maintained. However, if the number of charges within the {\it volume} of the unit sphere is of interest, then this function is inappropriate as the $[N^+]$ configuration contains $N+1$ charges and should be plotted to reflect this number.

The sawtooth fit function sketched in Fig.~\ref{fig:lafave-functions}b rises linearly from each $N$-charge energy solution to a new upper limit of each subsequent $N+1$ energy solution. If this linear rise is considered as a gradual ``charging up'' of $q_0$ at the origin of each $[N]$, then the slope of the function is $N/\Delta N = N$. The generally shorter drop from $U(N^+)$, (cf. Fig.~\ref{fig:lafave-Uplus}) plotted appropriately at each $N\!+\!1$, to $U(N+1)$ maintains the total number of charges in the system (including the spherical volume) as $q_0$ is displaced from the origin to the surface. During this transformation, the spatial symmetry of the system changes from that of $[N]$ to $[N+1]$. These discrete symmetry transformations are responsible for correlations made previously\cite{lafave-lafave2006, lafave-lafave2008, lafave-lafave2013} to electron shell-filling throughout the periodic table of elements as discussed in the next section.

Though not treated here, continuously smooth functions having periodic ($\Delta N = 1$) local minima, Fig.~\ref{fig:lafave-functions}c, or sharply defined (infinite derivative) behavior at integer values of $N$ as shown in Fig.~\ref{fig:lafave-functions}d may also be useful to obtain solutions of the Thomson Problem. Perhaps such a fit function may include a variety of polynomial series exhibiting oscillatory behavior. 

\section{Correspondence with Atomic Structure}
\begin{figure}%
\begin{center}
\includegraphics[width=7.5cm]{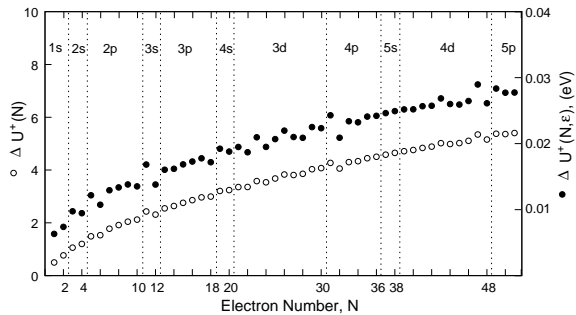}%
\caption{Discrete energy differences in the Thomson Problem, $\Delta U^+(N)$ (open circles) exhibit disparities uniquely correspondent to shell-filling behavior in natural atomic structure. These disparities are enunciated when the Thomson Problem is treated within a dielectric sphere (filled circles). These data are given for a dielectric sphere of 100nm radius and dielectric constant, $\varepsilon = 20\varepsilon_0$.}%
\label{fig:lafave-50}%
\end{center}
\end{figure}

The discrete derivative prohibits electrostatic energy ``states'' between integer values of $N$. Instead of beginning with an $N$-charge configuration we may begin with an $(N+1)$-charge configuration in Fig.~\ref{fig:lafave-symmetryDiagram}. The energy $\Delta U^+(N+1)$ represents an increase for the $(N+1)$ system by an amount approximately proportional to $\sqrt{N}$. The resulting $U(N^+)$ may then be associated with the $N$-charge configuration in the Thomson problem (using a strict ``surface-only'' interpretation). When plotted, Fig.~\ref{fig:lafave-50} (open circles), the distribution yields a pattern of energy differences, $\Delta U^+(N)$ (note the explicit association of energy with the $N$-charge configuration), consistent with electron shell-filling throughout the periodic table of elements as previously reported.\cite{lafave-lafave2006, lafave-lafave2008, lafave-lafave2013} In particular, as one progresses through the periodic table (increasing atomic number) the outermost electron in respective atomic structures is characterized by a wavefunction (or ``orbital'') having different spatial symmetries. For instance, the spherical $s$-shell associated with beryllium ($Z$=4) is followed by boron ($Z$=5), a system with an outermost electron having a dumbbell-shaped $p$-shell. These spatial symmetry differences among wavefunctions of neighboring atomic systems are repeatedly coincident with noticeable disparities found in $\Delta U^+(N)$ intrinsic to the Thomson Problem. Moreover, a relatively lower energy is obtained at the ``close'' of each electron shell followed by a disparity to a relatively higher energy which ``opens'' each subsequent shell. For example, a lower energy appears at $N$=4 followed by an abrupt disparity to a higher energy for $N$=5. 

The Thomson Problem treated within a dielectric sphere,\cite{lafave-lafave2006, lafave-lafave2008, lafave-lafave2013} using an appropriate model for discrete charges in the presence of dielectrics\cite{lafave-lafave2011} these disparities become more pronounced as shown in Fig.~\ref{fig:lafave-50} (solid circles). The shift, $N+1 \to N$, in numerical solutions of the Thomson Problem is unnecessary here. Note, for example, a dielectric sphere containing a single electron stores nonzero energy referenced to the empty, or electrically neutral, dielectric sphere. Enunciation of these features when using a dielectric sphere is partly due to the use of a ``Thomson radius'' as a variational parameter rather than a static unit sphere. Both frameworks may be traced to the classical ``plum-pudding'' model of the atom\cite{lafave-thomson1904} in which electrons reside within a uniform positively-charged sphere. A neutral dielectric sphere is used in place of the positively-charged sphere, and the Thomson Problem is a natural consequence of the removal of this positively-charged sphere in which electrons fly apart to form a spherical distribution.\cite{lafave-levin}

Many additional correspondences with atomic shell-filling have been reported.\cite{lafave-lafave2006, lafave-lafave2008, lafave-lafave2013}

\section{Concluding Remarks}
In place of coarsely-defined fit functions to numerical solutions of the Thomson Problem, the discrete derivative of these fit functions has been demonstrated to correspond with the well-defined configuration of $N$ charges on a unit sphere surrounding an $(N+1)$th charge, $q_0$, at its origin followed by relaxation of this volumetric problem to each subsequent $(N+1)$-charge solution of the Thomson Problem. This approach provides useful new information about the discrete charge system not obtained by the insufficient mathematical description of coarse fit functions previously used to model solutions of the Thomson Problem.

The existence of possibly better fit functions based on the underlying physical properties of the charge system described may lead to better approximations and potentially exact solutions of the Thomson Problem which appears to be intrinsically shared with atomic structure.

Discrete transformations between neighboring $[N]$ and $[N+1]$ configurations through an intermediate $[N^+]$ configuration utilizing $q_0$ at the origin include a trivial linear increase in energy followed by a non-trivial minimization of energy. The latter transformation from a well-defined configuration to the unknown $[N+1]$ configuration is the embodiment of the Thomson Problem. Further developments toward exact solutions of this transformation are merited, substantiated by its many correspondences with electron shell-filling in atomic structure, and its usefulness to a wide range of other discrete charge modeling applications. Identification of these discrete transformations in the Thomson Problem may be the first step toward the development of new and more efficient energy minimization algorithms.

\end{document}